\documentclass[12pt]{article}
\usepackage{amssymb, amsthm}
\usepackage{mathtools}
\usepackage{hyperref}
\usepackage{cleveref}
\usepackage{multicol}
\usepackage{graphicx}
\usepackage{float}
\usepackage{siunitx}
\usepackage[raggedright]{titlesec}
\usepackage[margin=1.0in]{geometry}
\graphicspath{ {./images/} }

\titlespacing*{\section}{0pt}{1.2\baselineskip}{0.6\baselineskip}
\titlespacing*{\subsection}{0pt}{1.2\baselineskip}{0.6\baselineskip}

\DeclareMathOperator{\LogAUC}{LogAUC}
\DeclareMathOperator{\rand}{rand}
\DeclareMathOperator{\EnrichmentScore}{EnrScore}  

\title{Enrichment Score: a better quantitative metric for evaluating the enrichment capacity of molecular docking models}
    \author{Ian Scott Knight\\ian.knight@ucsf.edu
\and Slava Naprienko\\naprienko@stanford.edu
\and John J. Irwin\\john.irwin@ucsf.edu}
\date{October 2022}

\begin{document}

\maketitle

\begin{abstract}
\centering\begin{minipage}{\dimexpr\paperwidth-3in}
    The standard quantitative metric for evaluating enrichment capacity known as \textit{LogAUC} depends on a cutoff parameter that controls what the minimum value of the log-scaled x-axis is. Unless this parameter is chosen carefully for a given ROC curve, one of the two following problems occurs: either (1) some fraction of the first inter-decoy intervals of the ROC curve are simply thrown away and do not contribute to the metric at all, or (2) the very first inter-decoy interval contributes too much to the metric at the expense of all following inter-decoy intervals. We fix this problem with LogAUC by showing a simple way to choose the cutoff parameter based on the number of decoys which forces the first inter-decoy interval to always have a stable, sensible contribution to the total value. Moreover, we introduce a normalized version of LogAUC known as \textit{enrichment score}, which (1) enforces stability by selecting the cutoff parameter in the manner described, (2) yields scores which are more intuitively meaningful, and (3) allows reliably accurate comparison of the enrichment capacities exhibited by different ROC curves, even those produced using different numbers of decoys. Finally, we demonstrate the advantage of enrichment score over unbalanced metrics using data from a real retrospective docking study performed using the program \textit{DOCK 3.7} on the target receptor TRYB1 included in the \textit{DUDE-Z} benchmark.
\end{minipage}
\end{abstract}

\bigskip

\begin{multicols}{2}

\section{Introduction}

In the field of computational drug discovery, the technique of molecular docking is widely used to predict how small molecules will interact with larger molecules (typically receptor proteins)\cite{key_topics_in_molecular_docking_from_2019}. The more favorable a given interaction is predicted to be, the higher the small molecule's expected binding affinity with the receptor. Generally, a higher predicted binding affinity is taken to mean that that small molecule is more likely to actually bind to that receptor. A small molecule that actually binds to a receptor is called a \textit{ligand}. 

A given docking program constitutes a docking model that estimates the binding affinity of candidate small molecules to a receptor of interest by performing a constrained minimization of the model's \textit{scoring function}\cite{key_topics_in_molecular_docking_from_2019}. Many scoring functions have been proposed\cite{key_topics_in_molecular_docking_from_2019}\cite{overview_of_search_algorithms_and_scoring_functions_from_2006}\cite{overview_of_scoring_functions_from_2006_1}\cite{overview_of_scoring_functions_from_2006_2}\cite{overview_of_scoring_functions_from_2019}\cite{docking_score_ml_1}\cite{docking_score_ml_2}, ranging from force field calculations to knowledge-based methods and even machine learning methods.

The quality of a docking model is measured by how well the model can distinguish ligands from non-ligands\cite{key_topics_in_molecular_docking_from_2019}\cite{practical_guide_to_lsd}. This is done by examining how the model scores a dataset of small molecules consisting of (1) known ligands (called \textit{actives}) and (2) molecules known or expected not to bind to the receptor (known as \textit{decoys}). The practice of using a dataset of actives and decoys to measure how well a docking model can distinguish between ligands and non-ligands is known as \textit{retrospective docking}\cite{dud}. This is the principal model validation technique available for docking models. Once properly validated, a docking model may be used to perform \textit{prospective docking}, which means scoring molecules of unknown activity\cite{dud}. A prospective docking screen usually involves the scoring of many millions or even billions of molecules\cite{practical_guide_to_lsd}.

Decoys can be generated for a given receptor in a number of ways\cite{practical_guide_to_lsd}\cite{decoy_generation_using_ml}, but there are also several established datasets of actives and decoys available as benchmarks\cite{dud}\cite{dude}\cite{dudez}. Decoy sets are often designed to be particularly difficult to discriminate from actives for certain targets, such that if a given docking model can successfully discriminate between them in the setting of retrospective docking then this warrants a stronger belief in the model's usefulness as a tool for enrichment in the setting of prospective docking\cite{practical_guide_to_lsd}.

An important goal in the field of computational drug discovery is to find a way to quantitatively measure the capacity of a given docking model to \textit{enrich} a set of molecules by reliably predicting mostly favorable interactions for actives and mostly unfavorable interactions for decoys\cite{overview_of_enrichment_metrics}. There are a number of enrichment metrics that have been developed\cite{overview_of_enrichment_metrics}\cite{evaluating_virtual_screening_methods}\cite{protocols_for_bridging_peptide_to_nonpeptide_gap}. For example, the metric known as \textit{enrichment factor (EF)} captures the idea of enrichment by equating it to the proportion of actives present in some top fraction of best scoring molecules\cite{enrichment_factor}. Quantitative metrics like EF also allow researchers to compare the performance of different docking models on the same dataset of actives and decoys\cite{evaluating_virtual_screening_methods}.

The metric known as \textit{LogAUC}\cite{rapid_context_dependent_ligand_desolvation} is one of the most popular metrics for evaluating the quality of molecular docking models\cite{dude}\cite{dudez}\cite{practical_guide_to_lsd}\cite{improving_docking_based_virtual_screening_ability}. However, it comes with a significant drawback: it depends on a cutoff parameter\cite{rapid_context_dependent_ligand_desolvation}\cite{managing_bias_in_roc_curves}. This cutoff parameter controls what the minimum value of the log-scaled x-axis is, which is a mandatory consequence of the choice to use the logarithmic scale. Unless this parameter is chosen carefully depending on the dataset, one of the two following situations occurs: either (1) some fraction of the first inter-decoy intervals of the ROC curve are simply thrown away and do not contribute to the metric at all, or (2) the very first inter-decoy interval contributes too much to the metric (even compared to the contribution of the second inter-decoy interval), and this comes at the expense of all inter-decoy intervals following the first interval.

We fix this problem with LogAUC by showing a simple way to choose the cutoff parameter based on the number of decoys which forces the first inter-decoy interval to always have a stable, sensible contribution to the total value. Moreover, we introduce a normalized version of LogAUC known as \textit{enrichment score}, which (1) enforces stability by selecting the cutoff parameter in the manner described, (2) is more intuitively meaningful, and (3) allows reliably accurate comparison of the enrichment capacities exhibited by different ROC curves.

\section{Receiver Operating Characteristic}
The statistical tool known as the receiver operating characteristic (AKA ``ROC curve'') is a plot of how the true positive rate and false positive rate of an indexed family of binary classifiers change as the index is varied. The ROC curve itself is a step function:

\[
    \operatorname{ROC}(x) = \begin{cases}\label{eq:roc}
    y_1, & \text{if $k_{0} < x \leq k_{1}$,} \\
    y_2, & \text{if $k_{1} < x \leq k_{2}$,} \\
    & \vdots \\
    y_n, & \text{if $k_{n-1} < x \leq k_{n}$,} \\
\end{cases}
\]

where $y_i$ are true positive rates and $k_i$ are false positive rates.

\begin{figure}[H]
\includegraphics[scale=0.17]{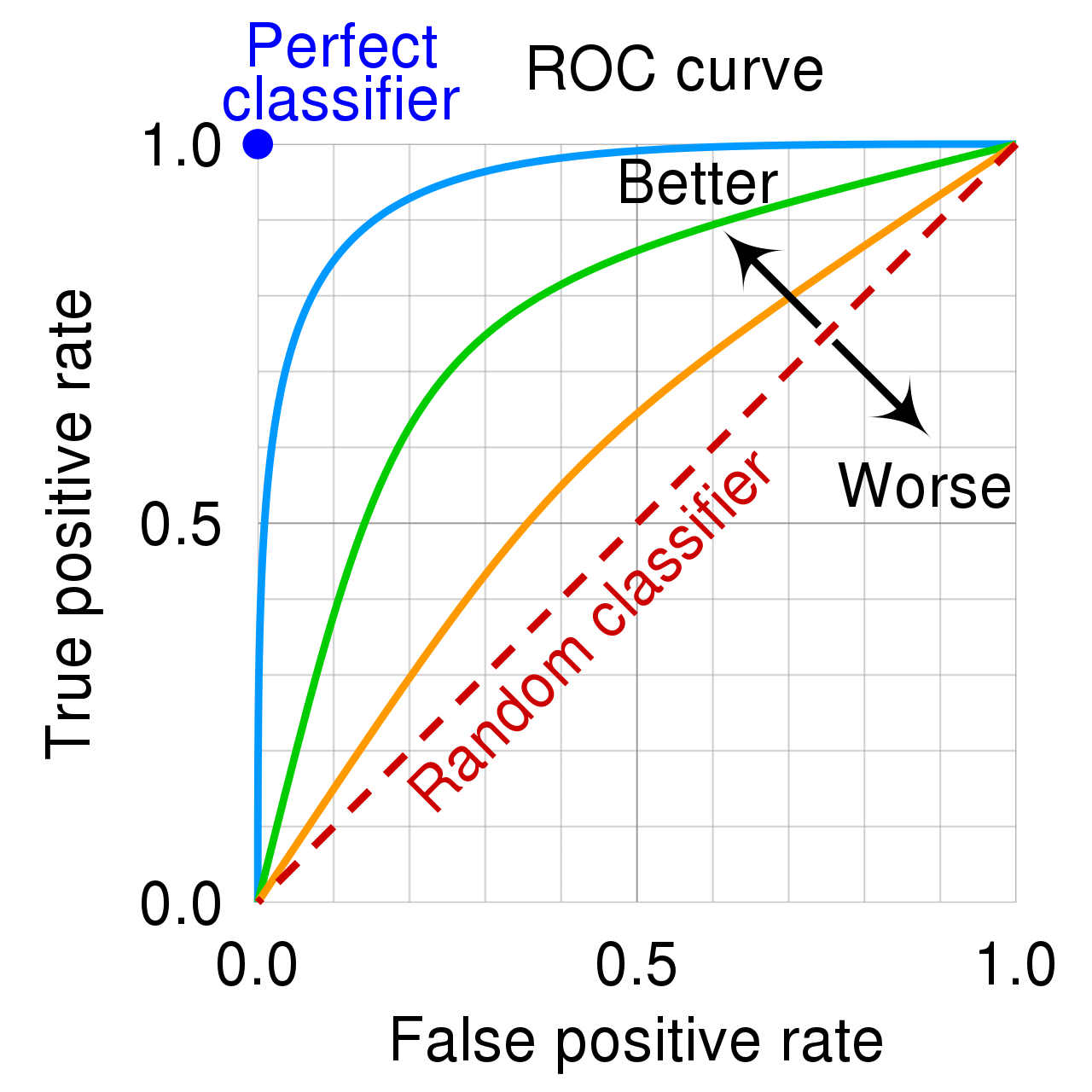}
\caption{An illustration of how to interpret the receiver operating characteristic.\cite{roc_curve_instructional_image}}
\end{figure}

To compute the ROC, we usually use the family of binary classifiers indexed on the real line where each particular classifier's index $i$ specifies a cutoff value below which it predicts $\operatorname{True}$ (``active'') and otherwise predicts $\operatorname{False}$ (``decoy''):

\[
    \text{classifier}_i(x) = \begin{cases}\label{eq:classifier}
        \text{True}, & \text{if $x < i$} \\
        \text{False}, & \text{if $x \geq i$}
    \end{cases}
\]

Now that we have our family of binary classifiers, we determine the true positive rate and false positive rate of each classifier in the family using the following dataset collected for a given set of molecules: (1) each molecule's docking model score $s_i \in \mathbb{R}$ and (2) its true active/decoy status $b_i \in \mathbb{B}$, where $\mathbb{B} = \{\operatorname{True},\operatorname{False}\}$ is the set of Booleans.

This description the family of binary classifiers typically used for computing ROC assumes that the docking model being used indicates more favorable predictions (i.e. those more likely to be ``active'') by assigning \textit{lower} scores. For example, if the docking model score is an approximation of the energy of the physical system formed by the scored molecule bound to the receptor, then a more negative score will indicate a more negative energy, which is more favorable. This is why the above classifier function predicts $\operatorname{True}$ for input below its index rather than above. If the docking model being used indicates more favorable predictions by assigning higher scores rather than lower scores, then the classifier function should be modified to instead predict $\operatorname{True}$ for input \textit{above} its index.

\section{Linear-log plot and LogAUC}

Calculated using a given docking model's predictions for a given set of molecules, the ROC curve is often used to analyze the trade-off between true positive rate and false positive rate in an indexed family of classifiers as its index is varied. While this visual presentation is useful, it is desirable to capture the enrichment capacity of a docking model in a single quantitative metric. The metric known as ``LogAUC'' is most commonly used for this purpose. In this section, we demonstrate the mathematical derivation of LogAUC and show how to calculate it explicitly for an ROC curve. 

First we define the literal logarithmic area under the curve for an arbitrary function $f$ in the form of an explicit integral. Let $f\colon [0, 1] \to [0,1]$ be an integrable function (any function you meet in practice is integrable). We consider the plot of $f$ with the $x$-axis on a logarithmic scale and the $y$-axis unchanged (on a linear scale). This kind of plot is known as a \textit{linear-log plot}. This logarithmic scaling of the $x$-axis is used in order to preferentially weight the early behavior of the function in determining the total value of LogAUC. In this way, LogAUC uses the well-understood mathematical operation of the logarithm to formalize the idea that earlier enrichment is significantly more important than later enrichment; in fact, because the logarithm is the inverse operation of exponentiation, the choice to use logarithmic scaling implies that earlier enrichment is, in particular, \textit{exponentially} more important than later enrichment.

Since $\ln(x)$ is undefined at $x = 0$, we need to decide an offset $0 < a < 1$ from zero to serve as the lowest value of $x$. This means we actually consider $f$ on $[a, 1]$. The function $f(x)$ on $[a,1]$ on a logarithmic scale turns into $f(e^x)$ on $[\ln(a), 0]$. (We remind that $\ln(a) < 0$ for $0 < a < 1$.) The area $A(f; a)$ of $f$ on a linear-log plot with the offset $a$ is given by

\begin{equation}\label{eq:logarea}
    \begin{aligned}
        A(f; a) &= \int_{\ln(a)}^{0}f(e^x)\,dx \\
    & = \int_a^1 f(y)\,\frac{dy}{y}.
    \end{aligned}
\end{equation}

The maximum area is attained when $f = 1$ on the entire interval $[0,1]$, giving $A_{\operatorname{max}} = A(f=1; a) = -\ln(a)$. 

The usual approach at this moment is to divide this area of function in the linear-log plot by its maximum possible area to get the quantity known as LogAUC:

\begin{align}\label{eq:logauc}
    \LogAUC(f; a) &= \frac{A(f; a)}{A(1; a)} \\
    & = \frac{1}{-\ln(a)}\int_a^1 f(x)\,\frac{dx}{x}.
\end{align}

For example, for the random classifier $\rand(x) = x$, we get 
\begin{align*}
    \LogAUC(\rand; a) &= \frac{1}{A(1; a)}\int_{a}^{1} x \, \frac{dx}{x} \\
    & = \frac{1-a}{-\ln(a)}.
\end{align*}

According to the literature \cite{rapid_context_dependent_ligand_desolvation, dude, dudez, practical_guide_to_lsd, improving_docking_based_virtual_screening_ability}, there is the standard offset value $a = \num{1.0e-3}$ that is typically used by practitioners who employ LogAUC to evaluate the enrichment capacity of their docking models.
    
If $a = \num{1.0e-3}$, then 
\begin{align*}
    \LogAUC(\rand; a) &= \frac{1-0.001}{\ln(1)-\ln(0.001)}\\
    &\approx 0.144622,
\end{align*}

which matches the value specified in the literature for the LogAUC of the random classifier\cite{dudez}.

But notice that the value $\LogAUC$ of the random classifier depends on the parameter $0 < a < 1$. For example, the LogAUC of the random classifier with the offset $a = 10^{-4}$ $\approx 0.108563$. It shows how unintuitive $\LogAUC$ is as a metric for enrichment capacity because (1) its value for the random classifier depends on the offset parameter, and (2) it does not provide an easy way to compare a given classifier against the random classifier. 

This makes it seem desirable to use a consistent value for $a$ across all datasets, so as to allow comparison of their LogAUC measurements.

\subsection{Dependence on the x-axis offset parameter $a$}
As has been noted elsewhere\cite{homology_modeling}, LogAUC measurements of different ROC curves are comparable only if they use the same cutoff parameter $a$. However, it is extremely important to point out that if $a$ is not set carefully for a given dataset, then there are two problems that can arise:

(1) $a$ is too high, resulting in some fraction of the lowest ranked molecules being simply discarded.

(2) $a$ is too low, resulting in the weight of the first term in the weighted sum becoming arbitrarily large, thereby causing the first term to dominate the score at the expense of the later terms. 

In both cases, some information in the dataset goes unused towards calculating LogAUC. In the former case, it is the information embodied in some number of the lowest ranked Booleans, and in the latter case, it is the information embodied in the higher ranked Booleans. In order to make use of all the information available in a given dataset, $a$ must be set in a principled way that avoids both of these problems.  

So, by using the same $a$ for all datasets, one still runs the risk of losing the ability to accurately compare different ROC curves against each other. Thus, in order to ensure that we may accurately compare different ROC curves, including those of different datasets, we need to set $a$ depending on the dataset in a way that ensures that $a$ is neither too large nor too small for that dataset.

To solve this problem, we propose setting the value of $a$ such that weight of the first term in the weighted sum is always 1. This is achieved by setting $a$ dynamically based on the number of decoys: $a = \frac{1}{en}$.

Suppose $f$ is a step function with values $y_i$ each on the interval $[\frac{i}{n}, \frac{i+1}{n}]$. Note that an ROC curve satisfies this definition. 

\[
    f(x) = \begin{cases}
        y_1, & \text{if $k_0 < x \leq k_1$} \\
        y_2, & \text{if $k_1 < x \leq k_2$} \\
        & \vdots \\
        y_n, & \text{if $k_{n-1} < x \leq k_n$} \\
    \end{cases}.
\]

Then we can compute the integral (\Cref{eq:logarea}) explicitly as a weighted sum of the values $y_i$ of the step function:
\begin{align}\label{eq:logareastep}
    A(f; a) &= y_1\int_{a}^{\frac{1}{n}}\frac{dx}{x} + \sum_{i=2}^{n}y_i \int_{\frac{i-1}{n}}^{\frac{i}{n}}\frac{dx}{x}.
\end{align}

Since an ROC curve is a step function, we use this weighted sum to calculate $A(\operatorname{ROC}; a)$.

Substituting our proposed value for $a$ into (\Cref{eq:logareastep}), we get:
\begin{align*}
    A(f; a=\frac{1}{en}) &= y_1\int_{\frac{1}{en}}^{\frac{1}{n}}\frac{dx}{x} + \sum_{i=2}^{n}y_i \int_{\frac{i-1}{n}}^{\frac{i}{n}}\frac{dx}{x} \\
    &= y_1 + \sum_{i=2}^{n}y_i \ln\left(\frac{i}{i-1}\right).
\end{align*} 

Let $n = 1000$. Then the terms in the sum look like
\[
    y_1 + 0.69 y_2 + 0.41 y_3 + 0.29 y_4 + \dots + 0.001 y_n
\]
Exactly as we want: the first term contributes the most, and the respective contributions of following terms diminish exponentially.

The last formula has obvious meaning: the first term has a weight of 1 by the choice of $a = \frac{1}{en}$ and all succeeding terms have exponentially diminishing weights, with the final term's weight $\ln(\frac{n}{n-1})$ coming arbitrarily close to zero with high enough $n$. Hence, the values $y_i$ matter less and less as $i$ increases. 
Moreover, setting $a$ in this manner forces all the weights to be invariant to $a$, which would not be the case otherwise due to the fluctuation of the first weight. Since the weight of the first term is always 1, we never have to worry about $a$ being too high or too low. 

\section{Enrichment Score}
Taking into account the discussion above, we propose an alternative metric called \textit{Enrichment Score} given by
\begin{equation}
    \resizebox{0.85\linewidth}{!}{%
        $\begin{aligned}
            \EnrichmentScore(f) = \frac{A(f; a=\frac{1}{en})-A(x; a=\frac{1}{en})}{A(1; a=\frac{1}{en})-A(x; a=\frac{1}{en})}
        \end{aligned}$%
    }.
\end{equation}

The enrichment score satisfies
\begin{enumerate}
    \item $\EnrichmentScore(f=\rand) = 0$,
    \item $\EnrichmentScore(f=1) = 1$,
    \item $\EnrichmentScore(f=0) < 0$,
    \item The weight of the first term contributes $1$, and the weights of subsequent terms diminish exponentially.
\end{enumerate}

Let us elaborate on why the enrichment score provides an intuitive understanding of the enrichment capacity of a given docking model.

First, the enrichment score of the random classifier $f=x$ is \textit{always zero}. It is intuitively appealing, as a random classifier merely shuffles data and should therefore be expected to provide zero enrichment.

If the enrichment score of a classifier is positive, then that means that the classifier performs better than the random classifier. If the score is negative, it performs worse than the random classifier. The higher the score, the better the classifier. 

The best classifier $f=1$ gives the maximum possible score, $1$. The value of the worst classifier $f=0$ varies depending on $a$, but it is \textit{always negative}. 

While the best and random classifier's enrichment scores are both invariant, it should be noted that the minimum value (i.e., the worst classifier's enrichment score) depends on $a$. However, it should also be the case that all models with negative enrichment scores are equally useless as tools for enrichment, so what the minimum score is should not matter in practice.

Let us compare the interpretability of the enrichment score to that of LogAUC with arbitrary parameter $a$. If $\LogAUC(f; a) = 0.12$, does it perform better or worse than the random classifier? One cannot tell. To get the answer, one needs to compute the LogAUC of the random classifier (which depends on $a$) and compare values, which is neither intuitive nor convenient.

We remark that the enrichment score is closely related to the LogAUC. To wit, the enrichment score is a \textit{normalization} of LogAUC involving a shifting (subtraction of random classifier LogAUC) and re-scaling (to enforce maximum enrichment score of 1). The relation is expressed by the following formula: 

\begin{equation*}
    \resizebox{1.0\linewidth}{!}{%
        $\begin{aligned}
            \EnrichmentScore(f) = \frac{\LogAUC(f; a=\frac{1}{en})-\LogAUC(\rand; a=\frac{1}{en})}{1-\LogAUC(\rand; a=\frac{1}{en})}
        \end{aligned}$%
    }
\end{equation*}

Therefore, one can still acquire an intuition of enrichment score by looking at a linear-log ROC plot. We emphasize that the enrichment score is \textit{not the area itself} but the normalized LogAUC. (On that note, it should be remembered that LogAUC itself is also not simply the log-scaled area under the curve, as shown in (\Cref{eq:logauc}).)

As one is more interested in a positive score (better than random classifier) than a negative score (worse than random classifier), it makes more sense to re-scale the score to induce the property that the maximum score (corresponding to the performance of the best classifier) is 1. And the careful choice of the offset parameter $a$ ensures that you don't throw away all of your data or inadequately weight the first term. 

\section{EnrichmentScore vs. unbalanced metrics: some examples}
In this section, we show how a poor choice of the offset parameter $a$ leads to one of the two scenarios previously described and demonstrate that using enrichment score avoids these problems.

For the first case, let us examine what happens when $a$ is too large. Consider an arbitrary ROC curve derived from a dataset with number of decoys $n=10000$. Taking $a = 1.0 \times 10^{-3}$, which is the standard value we find in the literature, we see clearly that the behavior of the ROC curve over the range $[0, 0.001)$ does not inform the enrichment score at all. This range accounts for the first 10 inter-decoy intervals: $[0.0001, 0.0002)$, $[0.0002, 0.0003)$, ..., $[0.0009, 0.001)$. So, using such a high value for $a$ means throwing away the information contained in those intervals. These discarded inter-decoy intervals can have at most the value of the first kept inter-decoy interval $y_{kept}$ (in this case, $y_{kept} = ROC(0.001)$). Throwing these intervals away allows the resultant enrichment score to become higher than it would be if these intervals were included in its calculation. This should be sufficient to demonstrate why having too high a value for $a$ is worth avoiding.

Next, let us examine what may happen when $a$ is too small for a given dataset. We consider a dataset formed by docking against one of the receptors in the \textit{DUDE-Z} dataset\cite{dudez}, a benchmark for molecular docking programs providing actives and property-matched decoys for 43 target receptors. Specifically, we docked against the target receptor labeled TRYB1, which is the protease known as tryptase beta-1. We used \textit{DOCK 3.7}\cite{dock_3.7} for our docking program. 

We performed two retrospective docking jobs on TRYB1, one with number of decoys $n = 100$ and the other with $n = 50$. In both cases, we use all 38 ligands included for TRYB1 in \textit{DUDE-Z} as actives. As for decoys, for the dataset defined by $n = 100$, we randomly selected 100 of the 1911 decoys included for TRYB1. In order to select the 50 decoys needed for the dataset defined by $n = 50$, we (1) docked on the dataset defined by $n = 100$, (2) sorted the decoys by docking score, and (3) deleted every other decoy in the sorted list (i.e., we removed the 2nd best scored decoy, 4th best scored decoy, ..., 100th best scored decoy). This procedure is intended to make the ROC curve for the dataset defined by $n = 50$ remain approximately the same shape as the ROC curve for the dataset defined by $n= 100$.

\begin{figure}[H]
\includegraphics[width=0.45\textwidth]{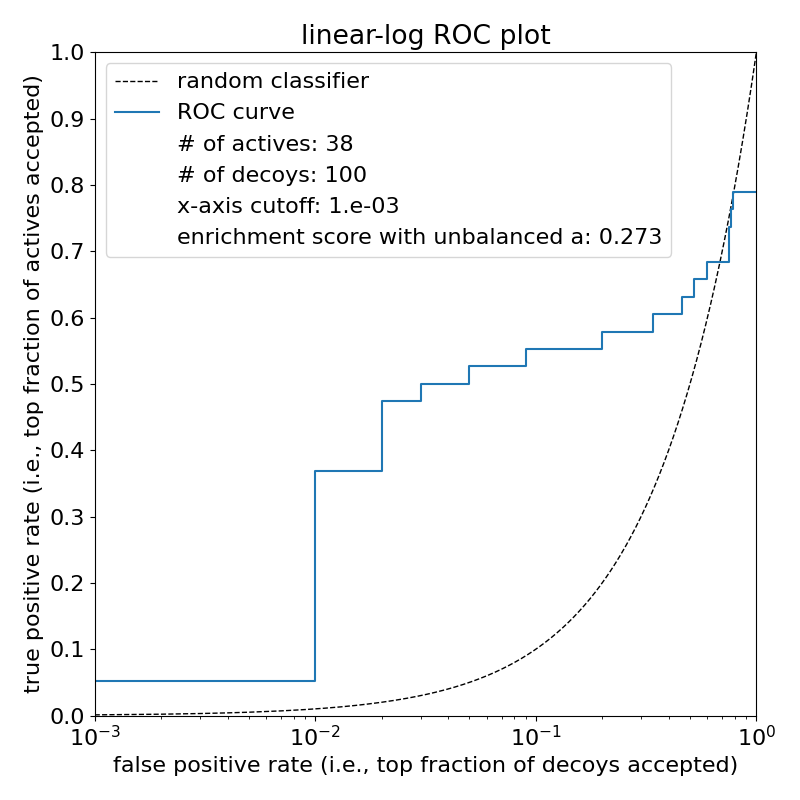}
\caption{ROC curve for TRYB1 using $n = 100$ and $a = \num{1.0e-3}$.}
\label{roc_tryb1_n=100_a=1.0e-3}
\end{figure}

\begin{figure}[H]
\includegraphics[width=0.45\textwidth]{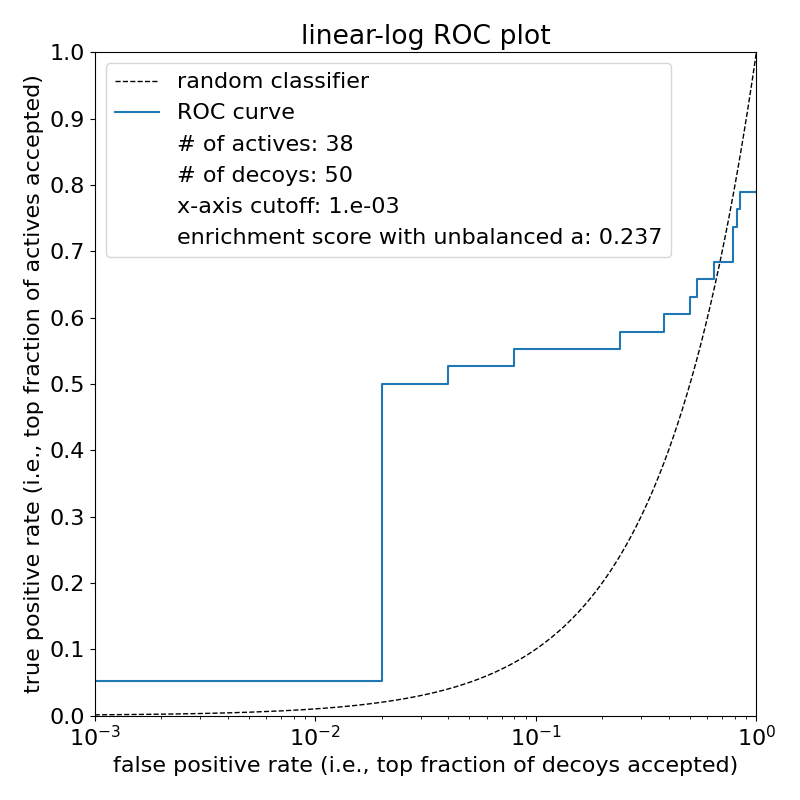}
\caption{ROC curve for TRYB1 using $n = 50$ and $a = \num{1.0e-3}$.}
\label{roc_tryb1_n=50_a=1.0e-3}
\end{figure}

Using the value $a = \num{1.0e-3}$ (previously mentioned to be the standard $a$ according to the literature), we get the following linear-log ROC plots for $n = 100$ (\Cref{roc_tryb1_n=100_a=1.0e-3}) and $n = 50$ (\Cref{roc_tryb1_n=50_a=1.0e-3}), respectively.

Notice that enrichment scores of the two plots vary, with a difference of 0.036.

\begin{figure}[H]
\includegraphics[width=0.45\textwidth]{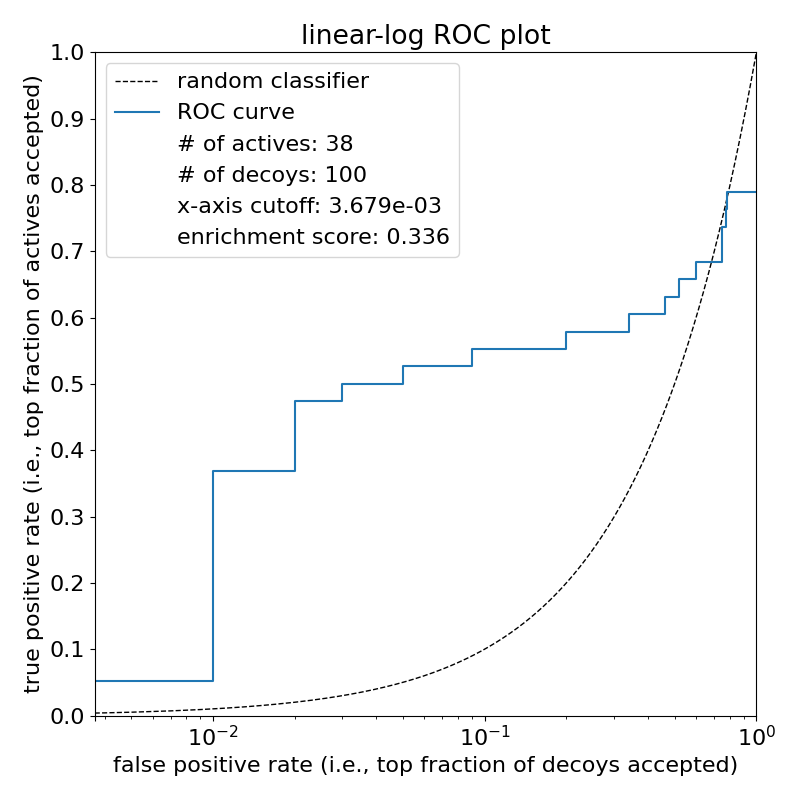}
\caption{ROC curve for TRYB1 using $n = 100$ and $a = \num{3.679e-3}$.}
\label{roc_tryb1_n=100_a=balanced}
\end{figure}

\begin{figure}[H]
\includegraphics[width=0.45\textwidth]{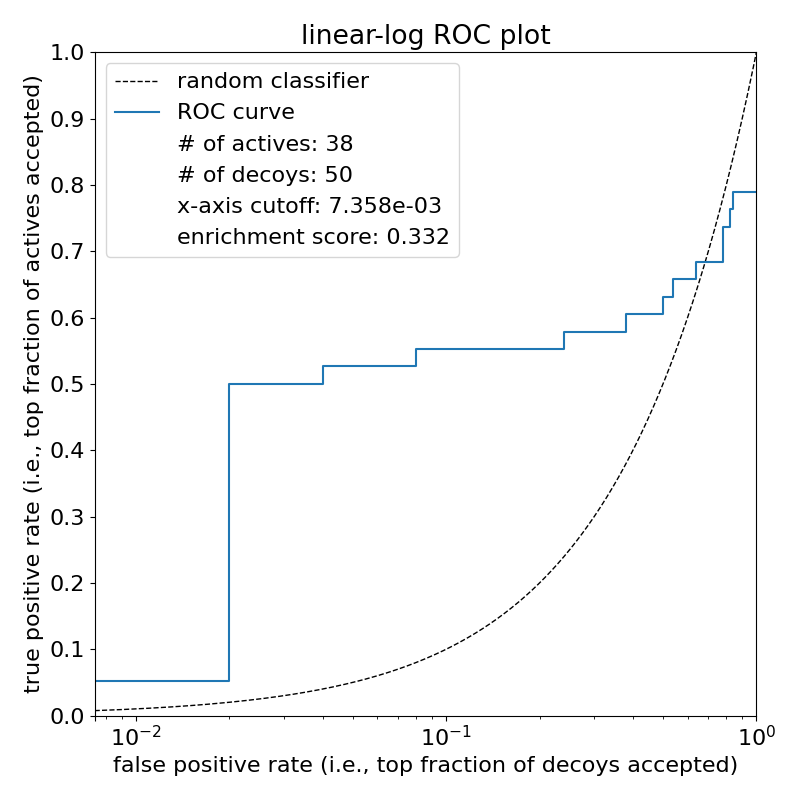}
\caption{ROC curve for TRYB1 using $n = 50$ and $a = \num{7.358e-3}$.}
\label{roc_tryb1_n=50_a=balanced}
\end{figure}

Now, let us plot the same data using $a = \frac{1}{en}$ in the following \Cref{roc_tryb1_n=100_a=balanced} and  
\Cref{roc_tryb1_n=50_a=balanced}.

\begin{figure}[H]
\includegraphics[width=0.45\textwidth]{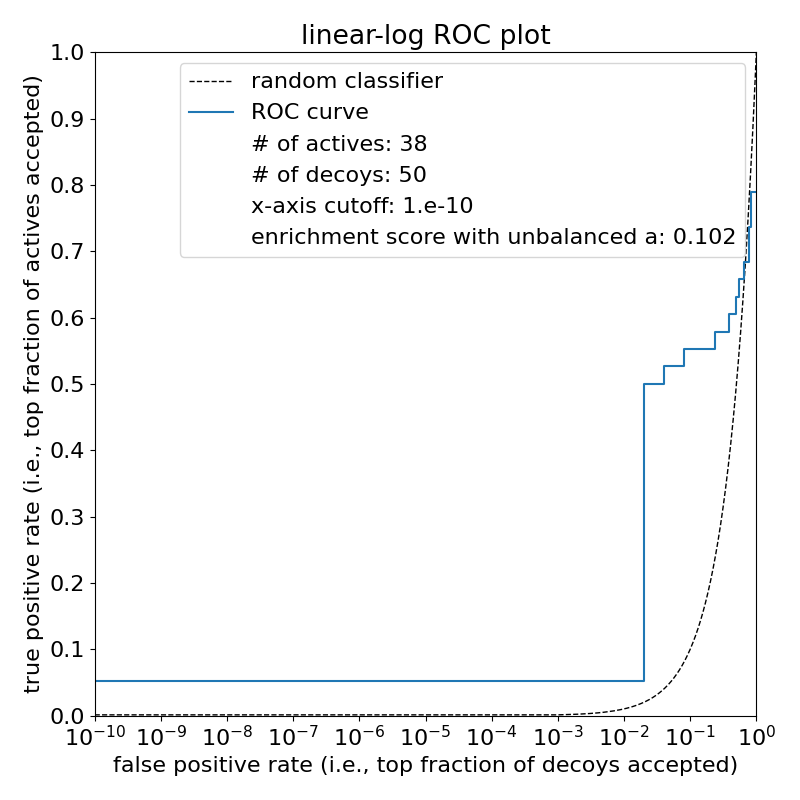}
\caption{ROC curve for TRYB1 using $n = 50$ and $a = \num{1.0e-10}$.}
\label{roc_tryb1_n=50_a=1.0e-10}
\end{figure}

$a = \num{3.679e-3}$ was used for $n = 100$, and $a = \num{7.358e-3}$ was used for $n = 50$. Notice that the enrichment scores for both are nearly identical, with a difference of only 0.004, which is considerably lower than the difference of 0.036 induced by using $a = \num{1.0e-3}$. This demonstrates that setting $a = \frac{1}{en}$ avoids the variability in enrichment score witnessed for different $n$ when using the apparently standard $a = \num{1.0e-3}$. Moreover, notice how far both enrichment scores produced using $a = \num{1.0e-3}$ are from those using $a = \frac{1}{en}$. For $n = 50$, the difference is a staggering 0.095. At this point, it should be noted that the difference in enrichment score using too low $a$ is determined by the value of the first interdecoy interval $y_1 = ROC(0)$, which the enrichment score comes arbitrarily close to with small enough $a$. Therefore, in the limit, having too low $a$ is tantamount to throwing away all inter-decoy intervals except the first one. The following figures (\cref{roc_tryb1_n=50_a=1.0e-4}, \cref{roc_tryb1_n=50_a=1.0e-5}, \cref{roc_tryb1_n=50_a=1.0e-10}) demonstrate this for the dataset defined by $n = 50$ using smaller and smaller $a$. 

\begin{figure}[H]
\includegraphics[width=0.45\textwidth]{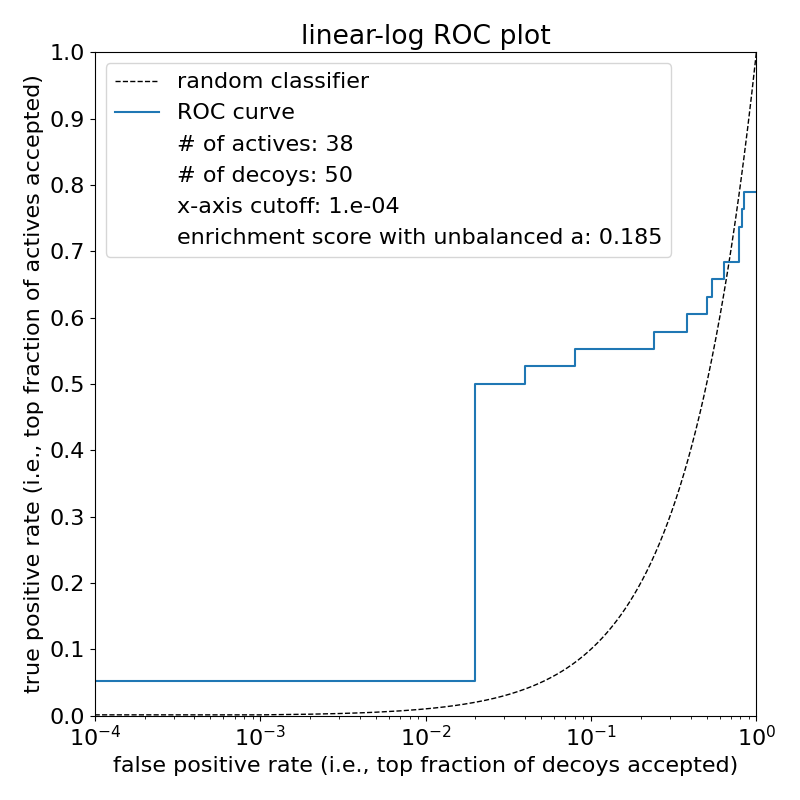}
\caption{ROC curve for TRYB1 using $n = 50$ and $a = \num{1.0e-4}$.}
\label{roc_tryb1_n=50_a=1.0e-4}
\end{figure}

\begin{figure}[H]
\includegraphics[width=0.45\textwidth]{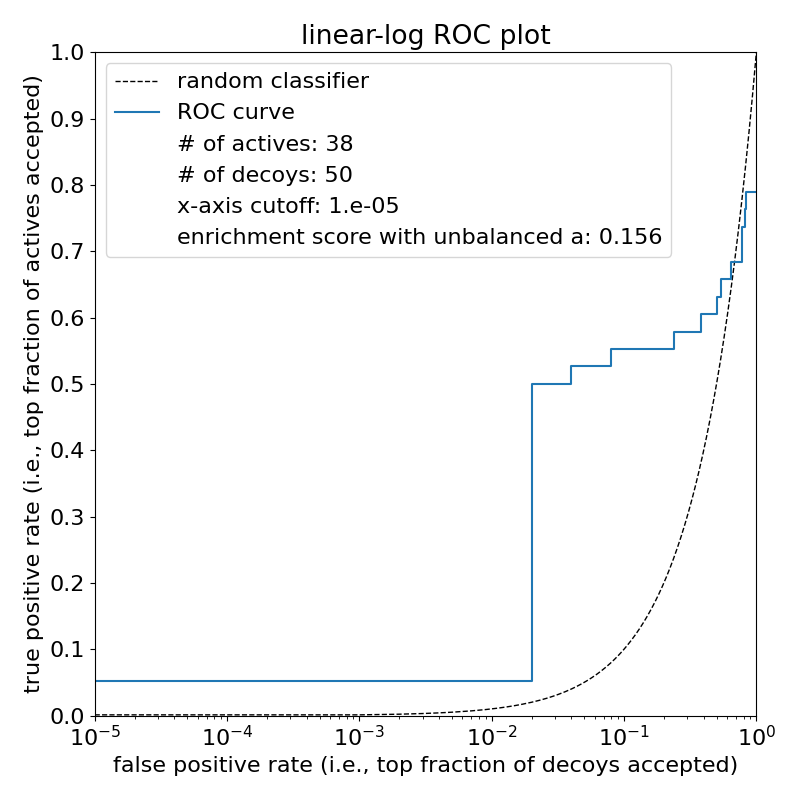}
\caption{ROC curve for TRYB1 using $n = 50$ and $a = \num{1.0e-5}$.}
\label{roc_tryb1_n=50_a=1.0e-5}
\end{figure}

We see now that the direction of error in enrichment score induced by a poor choice of $a$ is determined by the direction of error in choice of $a$: too high $a$ causes enrichment score to become higher than it should be, and too low $a$ causes enrichment score to become lower than it should be. In this way, the respective causes for the problems of using too high $a$ or too low $a$ complement each other.

\section{Data and Software Availability}

The software \textit{DOCK 3.7} is available through two licenses: (1) an academic and non-profit license, which is free for academic researchers and those engaging in non-profit research, and (2) a commercial license, which requires paying a license fee. More information is available at the \textit{DOCK 3.7} website: \url{https://dock.compbio.ucsf.edu/DOCK3.7/}. The \textit{DUDE-Z} dataset is publicly available at \url{https://dudez.docking.org/}.

\section{Funding}

J.J.I. was funded by \href{https://www.nigms.nih.gov/}{NIGMS} (Grant No. GM133836).

\end{multicols}

\end{document}